\documentclass{ws-ijmpa}

\usepackage{amsfonts,amsmath}
\usepackage{graphicx}
\usepackage{dcolumn}% Align table columns on decimal point
\usepackage{bm}% bold math

\begin{document}

\markboth{D. C. Latimer \& D. J. Ernst}
{Degeneracies of Mass-Squared Differences}

\catchline{}{}{}{}{}
%
%%%%%%%%%%%%%%%%%%%%%%%%%%%%%%%%%%%%%%%%%%%%%%%%%%%%%%%%%%%%%%%%%%%%

\title{ON THE DEGENERACIES OF THE MASS-SQUARED DIFFERENCES FOR THREE-NEUTRINO
OSCILLATIONS}

\author{D. C. LATIMER}

\address{
Department of Physics, University of Louisville\\
Louisville, Kentucky  40292, USA \\
dclati01@louisville.edu}

\author{D. J. ERNST}

\address{Department of Physics and Astronomy, Vanderbilt University\\
Nashville, Tennessee 37235, USA\\
david.j.ernst@vanderbilt.edu}

\maketitle

\begin{history}
\received{12 April 2005}
%\revised{Day Month Year}
\end{history}

\begin{abstract}
Using an algebraic formulation, we explore two well-known degeneracies involving the mass-squared
differences for three-neutrino oscillations assuming CP symmetry is conserved.  For vacuum oscillation, we 
derive the expression for the mixing angles that permit invariance  
under the interchange of two mass-squared differences.  
This symmetry is most easily expressed in terms of an ascending mass order.
This can be used to reduce the parameter space by one half in the absence of the MSW effect.
For oscillations in matter, we derive within our formalism the known approximate degeneracy between
the standard and inverted mass hierarchies in the limit of vanishing $\theta_{13}$.  This is done with a mass
ordering that permits the map $\Delta_{31} \mapsto -\Delta_{31}$.  
Our techniques allow us to 
translate mixing angles in this mass order convention into their values for the ascending order convention.
Using this dictionary, we demonstrate that the vacuum symmetry and the
approximate symmetry invoked for oscillations in matter are distinctly different.

\end{abstract}

\ccode{PACS numbers: 14.60.pq}

\section{Introduction}
In general, neutrino oscillations are achieved by relating, via a nontrivial unitary mixing matrix $U$,
flavor states to mass 
eigenstates of a Hamiltonian $H$; that is, $\nu_f = U \nu_m$.  
As the flavor states are not eigenstates of the Hamiltonian, flavor oscillation occurs
as the particles propagate through space.  The degree of mixing among the mass eigenstates is characterized
by the mixing matrix $U$; and the frequency of the oscillations is characterized by the differences in the squared
masses of these eigenstates.
Phenomenologists seek to determine the number of 
relevant neutrinos, the mixing matrix, and the mass-squared differences.  In a two neutrino theory, one
has one mass-squared difference, and
physical arguments reduce the mixing matrix to an element of the commutative group $U(1)$. As there are
only two parameters of interest, symmetries of the relevant formulae are readily apparent even if
one considers the additional complications brought on by matter effects.
For three neutrinos, the situation is more interesting but more complicated.  
The mixing matrix is now an element of a
noncommutative group $SU(3)$, parameterized by three real mixing angles and a Dirac phase; and one now must deal
with two independent mass-squared differences.  Adding matter effects into the description, further 
complicates the issue.  Though analytic expressions do exist in this general setting \cite{ohlsson}, 
any potential symmetries or degeneracies among the parameters are mired in awkward relations and
transcendental equations. For $N>3$ neutrinos, using traditionally formulated 
oscillation equations to determine symmetries is nearly hopeless even without matter effects.

In order to determine such degeneracies, one 
should consider neutrino oscillations in the context of the algebraic formulation 
developed in Refs.~\refcite{kchdcl1} and \refcite{kchdcl2}, valid for a fixed arbitrary number of flavors.  
We will restrict our discussion to a three neutrino theory; however, the methodology is
readily generalized.
In Ref.~\refcite{angles}, we utilized this formalism to
derive symmetry relations among the three mixing angles and CP phase in the mixing matrix. 
Here we shift our attention to symmetries involving the 
mass-squared differences.

We begin with a discussion of the discrete
CP and T symmetries for oscillation {\it in vacuo}.  The differences that we shall find between
the vacuum and matter oscillations are rooted in these symmetries.
Assuming CP invariance, we derive a 
degeneracy with respect to the mass-squared differences for three-neutrino vacuum oscillations.
This was observed by Ahluwalia \cite{dharam1}.  The symmetry is exact allowing one to
reduce the physically relevant parameter space by one half, as in the two-neutrino case.
In order to discuss the approximate symmetry for neutrino oscillations in matter, one must
appeal to phenomenology for guidance.  Data demonstrate that there are two distinct scales for the
mass-squared differences with one mixing angle near zero \cite{SS,list}.
We discuss this well-known approximate
degeneracy within our formalism.  
The two separate degeneracies are most easily accessed through different
mass ordering conventions.  We develop a dictionary so that we may relate
the two mass ordering conventions which allows us to demonstrate that these are two
distinct symmetries.

\section{Discrete symmetries}
We refer to Ref.~\refcite{kchdcl1} for a field-theoretic treatment of vacuum neutrino oscillations.  
The probability that an 
$\alpha$-flavor neutrino will be detected as a $\beta$-flavor neutrino at time $t$ after its creation is
\begin{equation}
\mathcal{P}_{\alpha \to \beta}(t,U) = \frac{1}{2} \mathrm{tr} [P_+ P^\alpha(t) P_+ P^\beta]. \label{osc1}
\end{equation}
The operators in this expression are defined as follows. The projection operator onto the 
flavor $\alpha$ is $P^\alpha$, with components in the mass basis $(P^\alpha)_{jk} = U^*_{\alpha j} U_{\alpha
k}$.
Time dependence
of the projection $P^\alpha (t)$ is as in the Heisenberg picture,
\begin{equation}
P^\alpha(t) = e^{i H t} P^\alpha e^{-i H t},
\end{equation}
where, in our case, we take $H$ to be a Dirac-like Hamiltonian.  The remaining operator $P_+$ (or $P_-$) projects
onto the positive-energy particle (negative-energy antiparticle) subspace,
\begin{equation}  
P_\pm= \frac{1}{2} \left( 1 \pm H\mathcal{E}^{-1} \right) ,
\label{ppm}
\end{equation}
where we define $\mathcal{E}$ to be the positive square root of the squared Hamiltonian.

The ability to swap the mass-squared differences is underpinned by the fundamental symmetries of time reversal 
and CP conjugation. We begin by explicating how these symmetries are implemented in the formalism.
As our Hamiltonian is Dirac, there exist
two antilinear involutions of import: time reversal, denoted by $T$, and CP conjugation, denoted by $\Sigma$.
Time reversal $T$ commutes with the Hamiltonian, and $\Sigma$ anticommutes with the Hamiltonian.
As $T$ is an involution and the trace is cyclic, we have from Eq.~(\ref{osc1})
\begin{equation}
\mathcal{P}_{\alpha \to \beta}(t,U) = \frac{1}{2} \mathrm{tr} [T P_+ P^\alpha(t) P_+ P^\beta T] \,.
\end{equation}
From the antilinearity of $T$, we achieve the equality
\begin{equation}
\mathcal{P}_{\alpha \to \beta}(t,U) = \mathcal{P}_{\alpha \to \beta}(-t, U^*)\,,  
 \label{tflip}
\end{equation}
where the dependence on $U$ is implicit in the definition of $P^\alpha$ as noted above.

The effect of the CP conjugation operator on the (anti-)particle
projection is
\begin{equation}
\Sigma P_\pm \Sigma = P_\mp\,,
\end{equation}
due to the anticommutativity of $\Sigma$ with $H$.
We also have
\begin{equation}
\Sigma e^{iHt} \Sigma = e^{iHt}\,,
\end{equation}
as $\Sigma$ is antilinear.
The consequence to the oscillation probability is 
\begin{eqnarray}
\mathcal{P}_{\alpha \to \beta}(t,U) &=& \frac{1}{2}\mathrm{tr} [\Sigma P_+ P^\alpha(t) P_+ P^\beta \Sigma] \\
&=& \mathcal{P}_{\bar \alpha \to  \bar \beta}(t, U^*)\,. \label{cpflip}
\end{eqnarray} 
From Eq.~(\ref{tflip}) and Eq.~(\ref{cpflip}), we note that the theory conserves CP and T separately if $U = U^*$. 
Considered jointly, we see that successive application of these operators indicates CPT invariance in general
\begin{equation}
\mathcal{P}_{\alpha \to \beta}(t,U) = \mathcal{P}_{\bar \alpha \to  \bar \beta}(-t, U).
\end{equation}
In what follows, we will restrict our study to a CP invariant theory; i.e., we set $U=U^*$.

\section{Vacuum oscillation}
In the ultra-relativistic limit ($E\gg m$), the oscillation probability Eq.~(\ref{osc1}) 
in vacuum for source-to-detector distance $L$ and
neutrino energy $E$ becomes
\begin{equation}
\mathcal{P}_{\alpha \to \beta}(L/E) = \mathrm{tr}
[e^{i\mathcal{M}L/2E} P^\alpha e^{-i\mathcal{M}L/2E} P^\beta]
\label{osctrace} 
\end{equation}
where $\mathcal{M}$, in the mass basis, is the diagonal matrix with entries $(m_1^2, m_2^2, m_3^2)$. 
In this formulation, it is
clear that adding multiples of the identity to $\mathcal{M}$ leaves the trace invariant; in other words,
it is only the mass-squared differences which are of dynamical consequence.  In this limit, time reversal invariance
is equivalent to invariance under the map $\mathcal{M} \mapsto -\mathcal{M}$. Making the summation
explicit,  the trace in Eq.~(\ref{osctrace}) is seen to yield the usual expression for the oscillation probability,
\begin{equation}
\mathcal{P}_{\alpha \to \beta}(L/E)
= \delta_{\alpha \beta}
-4 \sum^3_{\genfrac{}{}{0pt}{}{j >
k}{j,k=1}} U_{\alpha j} U_{\alpha k} U_{\beta k} 
U_{\beta j} \sin^2 (\varphi_{jk})
\label{oscform}
\end{equation}
where $\varphi_{jk} := \Delta_{jk} L/4E$ with $\Delta_{jk} := m_j^2 -
m_k^2$.  Here, the dependence on the mass-squared differences is readily apparent.  We remark 
that by definition they must satisfy
$\Delta_{12} + \Delta_{23} + \Delta_{31} = 0$;
hence, only two of the three are independent parameters.

In the context of vacuum oscillation, an exact degeneracy exists between these two independent mass-squared 
differences.  This degeneracy is valid for
all values of mass-squared differences and all mixing angles;
additionally, it is independent of the well-known approximate degeneracy for oscillations in matter.
This symmetry is most easily accessed if we insist upon the mass ordering
\begin{equation}
m_1 < m_2 < m_3. \label{ascendmass}
\end{equation}
Such an ordering is entirely general as permutations of the mass labels can be achieved via unitary transformations
which preserve the underlying physics.   Equivalently, we may characterize the oscillation dynamics with the
positive quantities $\Delta_{21}, \Delta_{32} >0$.  We make no assumptions regarding the relative values of these
two mass-squared differences, as our result holds in general.  
Derived herein, we may interchange the values of these two mass-squared
differences and maintain equivalent oscillation probabilities given a suitable change in the mixing matrix 
$U$.  As such, we need to adopt a parameterization of the mixing matrix; we use the standard 
representation \cite{pdg}
\begin{equation}
U(\theta_{23},\theta_{13},\theta_{12}) = \left(  
\begin{array}{ccc}
c_{12} c_{13} & s_{12} c_{13} & s_{13} \\
-s_{12}c_{23} - c_{12} s_{23} s_{13} & 
c_{12} c_{23} - s_{12} s_{23} s_{13}& s_{23}c_{13}\\
 s_{12}s_{23} - c_{12} c_{23} s_{13} & 
-c_{12} s_{23} - s_{12} c_{23} s_{13} & c_{23}c_{13}
\end{array}
\right) \,
\label{mixer}
\end{equation}
where $c_{jk} = \cos{\theta_{jk}}$,
$s_{jk}=\sin{\theta_{jk}}$, and the ranges on the mixing angles $\theta_{jk}$ are
discussed in Refs.~\refcite{gluza} and \refcite{angles}.
For our purposes, it will be useful to write the mixing matrix as three separate rotations 
\begin{equation}
U(\theta_{23},\theta_{13},\theta_{12})= 
\widetilde U_1(\theta_{23}) \widetilde U_2(\theta_{13}) \widetilde U_3(\theta_{12})\,.
\label{urot}
\end{equation}
Here, $\widetilde U_j(\theta)$ represents a proper rotation by angle $\theta$ about the $j$th axis in $\mathbb{R}^3$.

We are 
interested in the transformations of the mass-squared matrix which relabel the masses.
These are rotations about the $j$th axis  through an angle of $\pi/2$. The rotation
$\widetilde U_j(\pi/2)$ leaves the $j$th axis unchanged and interchanges the other two axes.
Conjugating $\mathcal{M}$ with $\widetilde U_2(\pi/2)$, for instance, yields
\begin{equation}
\mathcal{M}' := \widetilde U_2(\pi/2) \mathcal{M} \widetilde U_2(\pi/2)^\dagger = 
\mathrm{diag}(m_3^2, m_2^2, m_1^2)\,.
\label{2flip}
\end{equation} 
There are five nontrivial ways to permute the three masses; however, any of these
will disrupt the mass ordering adopted in Eq.~(\ref{ascendmass}).  In fact, we see that in the above example Eq.~(\ref{2flip})
the mass ordering is reversed, ${m_1'}^2 > {m_2'}^2 > {m_3'}^2$. 

We may invoke time reversal invariance, or
invariance under the map $\mathcal{M} \mapsto -\mathcal{M}$.  
This is simply a restatement that oscillation 
probabilities depend only on the absolute value of the mass-squared differences 
$\vert \Delta_{jk} \vert$, as is manifest in Eq.~(\ref{oscform}).  Given ascending masses
as in Eq.~(\ref{ascendmass}), the negative squared masses will satisfy the reverse inequality.  As such, if we effect the flip
of squared masses as in Eq.~(\ref{2flip}) and then change their sign, we have a new mass-squared matrix $\mathcal{M}'$ with
$\Delta_{21}' = \Delta_{32}$ and $\Delta_{32}' = \Delta_{21}$, which is what we desire.

The only remaining task is to determine what new mixing angles will result in 
an equivalent theory for mass-squared
matrices $\mathcal{M}$ and $\mathcal{M}'$.  Using the relation between the two in Eq.~(\ref{2flip}), we have
\begin{equation}
U\mathcal{M}U^\dagger = U\widetilde U_2(\pi/2)^\dagger\,\mathcal{M}' \,\widetilde U_2(\pi/2) U^\dagger\,.
\end{equation}
Clearly, the mixing matrix in question is $\widehat U =  U\widetilde U_2(\pi/2)^\dagger$.  To make this of practical use,
we must determine the new mixing angles $\theta_{jk}'$ in the chosen parameterization of Eq.~(\ref{mixer}).

In general, given the (real) matrix elements $U_{jk}$ of some $3 \times 3$ unitary matrix, 
we may parameterize the matrix in terms of the mixing
angles $\theta_{jk}$ as in Eq.~(\ref{mixer}) by solving the following transcendental equations
\begin{equation}
\tan {\theta_{12}}= U_{12}/ U_{11}, \quad \sin {\theta_{13}} = U_{13}, \quad \tan {\theta_{23}} = U_{23}/U_{33}
\end{equation}
assuming $U_{11}, U_{33} \ne 0$.  Should these matrix elements vanish, then we must deal with them
independently.  We first present these two special cases below and follow with the general situation.

Case (i):  $\sin{\theta_{13}}= 0$.  Since data analyses yield $\theta_{13}$ small \cite{SS,list}, 
this is a particularly interesting case. From the symmetry relations for the mixing
angles in Ref.~\refcite{angles}, 
it suffices to consider only $\theta_{13}=0$.
The new mixing matrix may be expressed as
\begin{eqnarray}
\widehat U &=& \widetilde U_1(\theta_{23}) \widetilde U_3(\theta_{12}) \widetilde U_2(\pi/2)^\dagger\nonumber\\
&=& \widetilde U_1(\theta_{23}+\pi/2) \widetilde U_2(\theta_{12}-\pi/2) \widetilde U_3(\pi/2);
\label{xx}
\end{eqnarray}
hence, the map is given by
\begin{equation}
\theta_{23}' = \theta_{23}+\pi/2, \qquad
 \theta_{13}'=\theta_{12}-\pi/2, \qquad  \theta_{12}'=  \pi/2. \label{vaccomp}
\end{equation}

Case (ii): $\cos{\theta_{12}}=0$. According to Ref.~\refcite{angles}, we need only consider
$\theta_{12}=\pi/2$.  We find
\begin{eqnarray}
\widehat U &=& \widetilde U_1(\theta_{23}) \widetilde U_2(\theta_{13}) \widetilde U_3(\pi/2) \widetilde U_2(\pi/2)^\dagger\nonumber\\
&=& \widetilde U_1(\theta_{23}+\pi/2)\widetilde U_2(0) \widetilde U_3(\pi/2-\theta_{13}),
\end{eqnarray}
so that
\begin{equation}
\theta_{23}' = \theta_{23}+\pi/2, \qquad
 \theta_{13}'=0, \qquad  \theta_{12}'=  \pi/2 -\theta_{13}.
\end{equation}

Case (iii):  The remaining situations. We
have
\begin{eqnarray}
\widehat U &=& \widetilde U_1(\theta_{23}) \widetilde U_2(\theta_{13}) \widetilde U_3(\theta_{12}) \widetilde U_2(\pi/2)^\dagger
\nonumber\\
&=& \widetilde U_1(\theta_{23}+\alpha) \widetilde U_2(\beta) \widetilde U_3(\gamma), 
\end{eqnarray}
where the angles satisfy the following set of transcendental equations
\begin{eqnarray}
\tan{\alpha} &=& \tan{\theta_{12}}\csc{\theta_{13}},  \nonumber \\
\sin{\beta} &=& -\cos{\theta_{12}} \cos{\theta_{13}}, \nonumber\\
\tan{\gamma} &=& \sin{\theta_{12}} \cot{\theta_{13}}. \label{gammavac}
\end{eqnarray}
The new angles are thus
\begin{equation}
\theta_{23}' = \theta_{23}+\alpha, \qquad
 \theta_{13}'=\beta, \qquad  \theta_{12}'=  \gamma.
\end{equation}

One consequence of this symmetry is that, in the context of vacuum oscillations, we may reduce the parameter
space as with the so-called ``dark side" in a two-neutrino theory.  One may choose to reduce the parameter space
by placing further constraints on the mass-squared differences, e.g., 
$\Delta_{21} < \Delta_{32}$.  Alternatively, one may look at all possible mass-squared differences while
further restricting the bounds on the mixing angles.

\section{Relating two mass order conventions}
This symmetry was readily realized for the convention of ascending masses; however, this choice is not the
dominant one found in the literature.  
Though our choice was made without loss of generality, we demonstrate the relation to the mass
ordering used extensively in three-neutrino phenomenology.
In order to account for the overwhelming majority of neutrino oscillation
data,  the two mass-squared differences need to differ by several orders of magnitude.  
Typically, one uses the convention
\begin{equation}
\Delta_{21} = \Delta_\odot > 0, \qquad \vert \Delta_{31} \vert = \Delta_\mathrm{atm} > 0,
\end{equation}
with $\Delta_\odot \ll \Delta_\mathrm{atm}$, cf. Refs. \refcite{SS} and \refcite{list}.  Though this convention establishes an ordering on the relative
absolute magnitudes of the mass-squared differences, it is entirely general in that it allows $\Delta_{31}$ to be
negative.  For $\Delta_{31}$ positive, one has what is termed the standard hierarchy.  This obeys the
ascending mass order that we previously adopted.  When $\Delta_{31}$ is negative, then one has an
inverted hierarchy.  Such can be achieved with the mass ordering
\begin{equation}
m_3 \ll m_1 < m_2.
\label{oorder}
\end{equation}
Although this convention has some distinct advantages, it can be conceptually difficult with which
to work \cite{gluza}. For the natural hierarchy, the ordering
of the mass eigenstates is by ascending mass. For the inverted hierarchy, the ordering is given
by Eq.~(\ref{oorder}), which constitutes a change of basis for one half of the problem.

For completeness sake, we derive the map between this formulation of the inverted hierarchy and the 
physically equivalent
one with ascending masses.  Suppose we have mixing angles $\theta_{jk}$ and mass-squared differences
$\Delta_{21} = \Delta_\odot$ and $\Delta_{31} = - \Delta_\mathrm{atm}$.  
Equivalent oscillation probabilities
can be got with the new mass-squared differences $\Delta_{21}''= \Delta_{\mathrm{atm}}$ and $\Delta_{32}'' =
\Delta_\odot$ (which obey the ordering $m_1'' < m_2'' < m_3''$ and $\Delta_{32}'' \ll \Delta_{21}''$) and mixing angles given by
\begin{eqnarray}
\theta_{23}''  =&  \theta_{23} - \arctan(1/\tan{\theta_{12} \sin{\theta_{13}}}), \nonumber\\
\theta_{13}''  =&  \arcsin(\sin{\theta_{12}} \cos{\theta_{13}}) ,  \nonumber \\
\theta_{12}'' =&  \arctan(\cos{\theta_{12}}/\tan{\theta_{13}}),  
\label{gammamsw}
\end{eqnarray}
whenever the operations are defined.  
Should $\sin{\theta_{12}} = 0$, then the new
mixing angles are
\begin{equation}
\theta_{23}'' = \theta_{23} - \pi/2 , \qquad \theta_{13}'' = 0, \qquad \theta_{12}'' =\theta_{13} - \pi/2\,. 
\end{equation}
Finally, the case of $\sin{\theta_{13}} = 0$, we have
\begin{equation}
\theta_{23}'' = \theta_{23} + \pi/2 , \qquad \theta_{13}'' = \theta_{12}, 
\qquad \theta_{12}'' = \pi/2\,.  \label{mswsym}
\end{equation}

\section{Oscillation in matter}
The inclusion of matter effects destroys the above symmetry.  
We shall give a brief exposition as to why this is the case and
then examine another well-known, but approximate, 
mass hierarchy degeneracy for the oscillation probability.

For oscillations in matter, we need to modify the Hamiltonian to account for the charged-current interaction
between the electron and electron-neutrino \cite{msw}.  To this end, one introduces an effective potential that 
operates only on the electron flavor.  At the level of the particle interaction, CPT is taken to be conserved;
however, as there exists a natural disparity between particle and antiparticle content of ordinary matter, 
one admits the possibility of extrinsic CPT
violation as discussed extensively in Ref.~\refcite{jacohl} and the references therein.  We continue to assume
an intrinsically CP invariant theory; however, the MSW effect will lead to an extrinsic violation
of this symmetry due to the change in sign on the effective potential for antineutrinos.
Oscillation probabilities for an oriented path $\Gamma$ through the matter are given by \cite{angles}
\begin{eqnarray}
\mathcal{P}_{\alpha \to \beta}(\Gamma,E) &=& \mathrm{tr} \left[
\exp \left\{ i \int_\Gamma \widetilde{H} dx \right\}  P^\alpha \right. \nonumber \\
&& \left. \exp \left\{ -i \int_\Gamma  \widetilde{H}  dx \right\} P^\beta \right]. \label{oscmass}
\end{eqnarray}
where, in the ultra-relativistic limit and modulo a multiple of the identity, the argument of the path ordered
exponentials in this trace can be 
written as
\begin{equation}
\int_\Gamma \widetilde{H}  dx = \mathcal{M} L/2E + 
\int_\Gamma A(x) dx P^e.  \label{integral}
\end{equation}

For antineutrinos, one needs to change the sign on the potential $A(x)$.  This leads to an extrinsic
violation of CP symmetry even though we have $U=U^*$.  As such, we are unable to reverse the ordering
of the masses via the map $\mathcal{M} \mapsto -\mathcal{M}$, a crucial step in the previous
derivation.  However, as is known, there exists another approximate symmetry between the regular and inverted
mass hierarchies.  

The presence of the effective potential complicates our cause, but not entirely.
If there exists some nontrivial subspace on which 
the MSW potential commutes with
$\mathcal{M}$, then we say that the
MSW effect has decoupled on that subspace.  Physically, the MSW effect does seem to decouple,
at least, approximately.  Let us write the mass-squared matrix as 
$\mathcal{M} = \mathrm{diag}(0, \Delta_{21}, \Delta_{31})$.
In the ideal limit of vanishing $\theta_{13}$, one notes that the image of the electron
flavor projection $P^e$ lies wholly within the subspace spanned by the $m_1$ and $m_2$ mass eigenstates.
We denote by $P_3$ the projection onto the eigenstates of mass $m_3$.  The MSW potential commutes
with this projection.  Using orthogonal projections, 
we separate the integral in Eq.~(\ref{integral}) into two commuting terms
\begin{equation}
\int_\Gamma \widetilde{H}  dx = (1-P_3) \int_\Gamma \widetilde{H}  dx  + P_3 \int_\Gamma \widetilde{H}  dx;
\end{equation}
to condense notation, we shall call the first operator on the RHS $B$ and the second $C$.
The second term is decoupled from the matter potential
\begin{equation}
C = P_3 \int_\Gamma \widetilde{H}  dx = P_3 \mathcal{M} L/2E
\end{equation}
where $P_3 \mathcal{M} = \mathrm{diag}(0, 0, \Delta_{31})$.
As $B$ and $C$ commute, the following applies $\exp\{i(B+C )\} = \exp(iB) \exp(iC)$.
In this limit, we consider the electron-electron oscillation probability in matter; that is, we set $\alpha=\beta=e$ 
in Eq.~(\ref{oscmass}).  Clearly, $\exp(iC)$ commutes with the flavor observable $P^e$ so that the operators involving
$\Delta_{31}$ cancel each other, making the probability independent of this parameter.
  
It is this independence that prompts us to consider negative values for $\Delta_{31}$, while leaving unchanged
$\Delta_{21}$ so as to not disrupt the MSW effect.  As shown, the change of sign of $\Delta_{31}$ does not affect the $\nu_e-\nu_e$
oscillation probability in matter but it does yield the inverted hierarchy.
We discuss the ramifications for the other probabilities, in particular, the vacuum oscillations.
The first consideration is the effect upon the remaining mass-squared difference.  Under the map, we have
$\Delta_{32} \mapsto -\Delta_{31} -\Delta_{21}$. 
In the relevant limit of $\Delta_{21} \ll \vert \Delta_{31} \vert$, 
one has to a good approximation $\Delta_{32} \mapsto -\Delta_{32}$.
The vacuum oscillation formula, Eq.~(\ref{oscform}), 
demonstrates that the oscillations will remain unchanged upon these two changes in sign.
As such, we have an approximate symmetry provided $\theta_{13}$ is near zero
and $\Delta_{21} \ll \vert \Delta_{31} \vert$.

We can use the results in Eq.~(\ref{gammamsw}) to determine what this approximate symmetry looks like
in the convention of ascending masses.  Though they are of intrinsic interest, these relations
allow us to compare the near degeneracy between the standard and inverted mass hierarchy in matter
with the exact vacuum symmetry relating the two hierarchies.
For ease of comparison, we restate
the approximate symmetry for the inverted mass hierarchy with matter effects in the ascending mass convention,  
\begin{equation}
\theta_{23}'' = \theta_{23} +\pi/2 , \qquad \theta_{13}'' = \theta_{12}, 
\qquad \theta_{12}'' = \pi/2\,.  
\end{equation}
Likewise, the vacuum symmetry for $\theta_{13}=0$ as discussed in Case (i), Eq.~(\ref{vaccomp}), is
\begin{equation}
\theta_{23}' = \theta_{23} +\pi/2, \qquad
 \theta_{13}'=\theta_{12} - \pi/2, \qquad  \theta_{12}'=  \pi/2. 
\end{equation}
Putting aside the correction to the mass-squared difference $\Delta_{32}$, we note
the difference between the mixing angles $\theta_{13}''$ and $\theta_{13}'$ for the two degeneracies.  
This clearly demonstrates that the approximate degeneracy between the two hierarchies for neutrinos
propagating in matter is {\it not} a mere perturbation of the exact vacuum symmetry.

We examine the matter mass hierarchy for a realistic case. We use a model \cite{us} 
to fit oscillation parameters to data. The results yield a
value of $\theta_{13} = 0.13$, a small but not insignificant number. The matter hierarchy has so
far been derived here neglecting terms of order $\Delta_{21}/\Delta_{31}$. We define the symmetry so as to
preserve the spacing between the levels,
\begin{equation}
\Delta'_{31} = -(\Delta_{31}-\Delta_{21})\,.
\end{equation}
A numerical calculation yields that with this definition, the resulting oscillation
parameters remain unchanged to the order of several tenths of a percent, even though $\theta_{13}$ is
0.13. A numerical refit of the data yields physical results that are even more similar  between the hierarchies.

\section{Conclusion}
We have examined, within our algebraic formalism, well-known symmetries which relate to the mass hierarchy question.
We work in two different conventions for the mass ordering for the inverted hierarchy and provide the
dictionary that relates the two conventions. We first investigate a symmetry which, in the absence of
matter effects, is exact for all values of the parameters. 
We find this symmetry reduces the allowed parameter space by one-half in the
absence of matter effects, just as occurs in the two neutrino case. We also derive
the mass hierarchy symmetry which holds in the presence of the MSW effect 
in the limit of small $\theta_{13}$ and two different scales for
the mass-squared differences. Utilizing our dictionary that translates between the two conventions for the
mass ordering, we show that these two symmetries are, in fact, different. 
For vanishing $\theta_{13}$, the relations for the mixing angles effecting the vacuum symmetry are given in
Eq.\ (\ref{vaccomp}) while the relations for the approximate symmetry in matter are in Eq.\ (\ref{mswsym}).
Finally, we find, like others \cite{jacohl},
that even for modest values of $\theta_{13}$ the approximate mass hierarchy symmetry is quite accurate. Given that
the present ability to perform neutrino oscillation experiments yields errors substantially larger
than the difference between the two hierarchy solutions, one tends to think of them as equivalent
solutions when phenomenologically extracting mixing angles and mass squared differences from data. The differences 
between the mass hierarchies are important in the context of looking for CP or CPT violation, themselves very small effects. However, there is other physics which 
in principle distinguishes between these solutions. An example would be gravitationally induced neutrino
oscillations \cite{dva} or oscillation in high density systems such as supernovae. 

\bibliography{massesb}
\end{document}